# Effect of radially heterogeneous band gap collapse on formation of swift heavy ion tracks in $Al_2O_3$

R. Voronkov[1], D. Zainutdinov[1], N. Medvedev[2,3], A.E. Volkov[1]

*[1] P.N. Lebedev Physical Institute of the Russian Academy of Sciences, Leninskij pr., 53,119991 Moscow, Russia,* roman.a.voronkov@gmail.com

*[2]Institute of Physics, Czech Academy of Sciences, Na Slovance 2, 182 21 Prague 8, Czech Republic;*

*[3]Institute of Plasma Physics, Czech Academy of Sciences, Za Slovankou 3, 182 00 Prague 8, Czech Republic;*

## Abstract

We estimate the effects of radial heterogeneity in the collapse of the electronic band gap on the damage in $Al_2O_3$ after impact of a swift heavy ion decelerated in the electronic stopping regime. The Monte Carlo code TREKIS describes the initial excitation of the electronic and ionic systems following the ion passage, while the density functional theory based molecular dynamics traces changes in the band structure in the ion track. This combination of methods enables us to compute the profile of energy transferred to the lattice by the time of relaxation of the electronic excitation, accounting for the induced spatial inhomogeneity of the band structure around the ion trajectory. We demonstrate that impact of a 700 MeV Bi ion induces a transient metal-semiconductor heterojunction in $Al_2O_3$: the metallization (the band gap collapse) occurs within a radius of about 2 nm from the ion trajectory. The band gap shrinks at distances of about 3-5 nm, while it remains almost unaffected at radii larger than 5 nm. Using this data, we estimate the atomic heating depending on the degree of band gap reduction at different radii from the ion trajectory. This approach refines the damage modeling, producing more pronounced discontinuous damage patterns along the ion path for all crystallographic directions compared to the model that assumes all the energy accumulated in the electron-hole ensemble is delivered to the atoms.

## Introduction

Swift heavy ions (SHIs, with energies $E$ > 1 MeV/nucleon and masses $M > 20\ m_p$, where $m_p$ is the proton mass) deposit the majority of their energy into the electronic system of a target [1,2]. The atomic response to the energy transferred from the excited electronic system results in unusual nanometric material modifications along the ion trajectory [3]. This property makes SHIs a valuable tool for various applications, from material nanostructuring to radiation cancer therapy [4–6], which require reliable and predictive theoretical models to facilitate the development of experiments and technology.

Ref. [4] presented the multiscale model that effectively describes the damage induced by SHIs in various materials. The model combines the original Monte Carlo code TREKIS-3 with molecular dynamics (MD). TREKIS traces the excitation of the electronic and atomic systems of a target after an SHI impact and provides the radial distribution of the energy transferred to target atoms, which is subsequently used as an initial condition for MD simulations of the atomic system's response to the excitation [5].

The application of this approach established that the transfer of the potential energy of generated electron-hole pairs into the atomic system is an important and extremely fast (shorter than 50-100 fs) mechanism for the excitation of insulators and semiconductors around the SHI trajectory [6]. It was shown that the driving force of this transfer is nonthermal band gap collapse (changes in the electronic band structure) caused by fast atomic displacements due to sharp changes in the interatomic potential energy surface under extreme electronic excitation [7,8]. The following assumptions were made introducing this nonthermal mechanism to the description of damage formation in SHI tracks in a number of materials [9–12]: (a) the band gap is assumed to collapse within ~100 fs (a typical timescale of the relaxation of the electronic system in a track) in the region excited after an SHI passage, and (b) all the potential energy of electron-hole pairs is transferred to the atomic system of the target.

However, these approximations need additional scrutiny. Indeed, their applicability to various materials and conditions remains underexplored.

First, a material's vulnerability to nonthermal band gap collapse strongly depends on bond ionicity [13]. For example, in alkali halides, the band gap does not collapse even at extremely high deposited energy densities (up to 20 eV/atom), although a portion of the electron-hole pairs' potential energy can be transferred to atoms due to band gap shrinkage [13].

Second, the radial distribution of the electronic excitations is spatially non-uniform within an ion track and quickly decreases from the ion trajectory outwards [14]. At a certain radius from the track center (which is yet to be estimated), the density of the energy deposited into the electronic system becomes insufficient for the initiation of the nonthermal transitions. Thus, the band gap does not completely collapse in the entire excitation region, and only a portion of the potential energy of generated electron-hole pairs' transfers to the atomic lattice via the above-mentioned mechanism.

On the other hand, scattering of excited electrons simultaneously heats up the atomic system, which may accelerate nonthermal processes in the excited matter in an SHI track [15]. At the same time, the electronic system loses energy during this scattering, which may hinder the nonthermal effects.

Summarizing, it is *a priori* unclear whether the energy deposited into the electronic system of a particular target around the SHI trajectory is sufficient for the band gap collapse or not. Thus, an accurate calculation of the nonthermal effects on the SHI track formation in an arbitrary material requires self-consistent modelling that includes simulation of both, electron scattering and nonthermal effects. The latter can be described within *ab-initio* methods such as the density functional theory (DFT) or the tight-binding method (TB). However, a direct *ab-initio* simulation of track formation is challenging due to the large track size compared to the typical *ab-initio* cell sizes, as well as high energy gradients.

By employing a simplified method of accounting for nonthermal effects, this paper lays the groundwork for the further development of a multiscale model of SHI track formation, providing qualitative insights into the effect of spatially nonuniform band gap collapse on the kinetics in SHI tracks. It is illustrated on the example of $Al_2O_3$ as a material well studied both experimentally and theoretically [16–18].

## Model

We study the problem of spatially inhomogeneous band gap collapse via the combination of methods.

In the first step, we use the Monte-Carlo (MC) code TREKIS-3 [19] for description of the passage of an SHI, development of subsequent electronic cascades, ionization of matter, formation and transport of secondary electrons and holes, as well as energy transfer to the material's atomic lattice through electrons and holes scattering on the lattice [14,20]. It also describes the Auger decay of core holes and the transport of photons produced in the radiative decays.

TREKIS-3 uses the asymptotic trajectory method with the scattering cross sections that take into account collective response of the target electronic and ionic systems to excitation. It relies on the framework of the dynamic structure factor formalism expressed through the complex dielectric function [21,22]:

$$\frac{d^2\sigma_{e,at}}{d(\hbar q)d(\hbar\omega)} = \frac{\left[Z_{eff}(v)Z_t(v)e\right]^2 m_p}{\hbar^2\pi E n_{at}} \frac{1}{\hbar q}\left(1 - e^{-\frac{\hbar\omega}{k_b T_{irr}}}\right)^{-1} Im\left[-\frac{1}{\varepsilon_{e,at}(\omega,q)}\right] \tag{1}$$

Here the indices *e* and *at* denote the scatterings on electrons or atoms, respectively; $n_{at}$ is the atomic concentration; $\hbar\omega$ and $\hbar q$ are the energy and momentum transferred from the projectile to the target, and $\hbar$ is the Planck's constant; *E* is the energy of the projectile and $m_p$ is its mass. The valence hole mass is determined within one-band model based on the density of states of the valence band [20,23]. $Im\left[-\varepsilon_{e,at}^{-1}(\omega,q)\right]$ is the electronic or atomic (phononic) part of the loss function of the target (Ref.[14] presents the loss function for $Al_2O_3$); $T_{irr}$ is the irradiation temperature; *e* is the electron charge; $Z_{eff}$ is the effective charge of the incident particle in units of the electron charge ($Z_{eff}(v) = 1$ for electrons and valence holes, whereas, the Barkas formula is used for an SHI [25]). $Z_t$ is the charge of the scattering centers.

For target electrons, $Z_t$=1. For atoms, the charge $Z_t(v)$ is calculated according to the algorithm described in Ref. [24]. This algorithm accounts for the dynamical screening effect, decreasing the screening of the atomic nucleus with the increase of the velocity of the incident projectile. For example, the nucleus is completely unscreened when the velocities of valence and inner-shell electrons are significantly smaller than the velocity of the incident particle: the atomic electrons have no time to adjust to the field generated by the projectile.

The results of the MC simulations include the temporal evolution of radial distributions of electron and valence hole densities around the SHI trajectory, their energy, and the density of the energy transferred to the atomic lattice. For statistical reliability, all results were averaged over $10^3$ MC iterations.

To estimate the likelihood of band gap collapse at varying distances from the ion trajectory, we divide the track area into 1-nm-thick cylindrical layers. Using the TREKIS output, we then determine the spatially and temporally averaged electronic and atomic kinetic temperatures within each layer. The results are averaged over the times from 10 fs to 100 fs after the SHI impact, providing an estimate of transient characteristic values of the temperatures in SHI tracks (excluding the initial highly nonequilibrium stage of Auger and electron cascades). Then, using the obtained characteristic temperatures, we apply the Quantum Espresso code to perform DFT-MD calculations and trace changes in the band structure and atomic temperature in each layer in $Al_2O_3$ [26].

The simulation box for MD modeling is initially equilibrated as follows. After the steepest descent geometry optimization, the initial kinetic temperature of the atomic lattice (proportional to the average kinetic energy) is set at $T^i_{kin}$ = 600 K by assigning atomic velocities according to the Gaussian distribution. Then, the atomic ensemble is equilibrated via DFT-based molecular dynamics for 500 fs, until the atomic system temperature reduces to oscillations around 300 K, according to the virial theorem [27]. The resulting atomic positions are then used as initial structures for further calculations, in which the lattice kinetic energy is increased by setting the kinetic temperature $T^i_{kin}$ = 300 K + $\Delta T^i_{kin}$, where $\Delta T^i_{kin}$ is the additional kinetic energy transferred to the atoms via the scattering of electrons and valence holes calculated with the TREKIS code.

In an SHI track, the electronic system follows the so-called bump-on-hot-tail distribution with the majority of slow electrons being in thermal equilibrium (following the Fermi-Dirac distribution) and a minority of fast ballistic electrons [28]. We set in the DFT calculations the electronic temperature that yields the same excess electronic energy per atom as that obtained from TREKIS simulations (which includes electrons' and holes' kinetic energy as well as the potential energy of electron-hole pairs).

To determine the dependence of the electronic temperature used in DFT calculations on the electronic energy, i.e. to estimate the electronic specific heat capacity, we carried out a series of self-consistent DFT calculations at various electronic temperatures. Figure 1 shows the dependence of the energy of the electronic system in alumina on the electronic temperature.

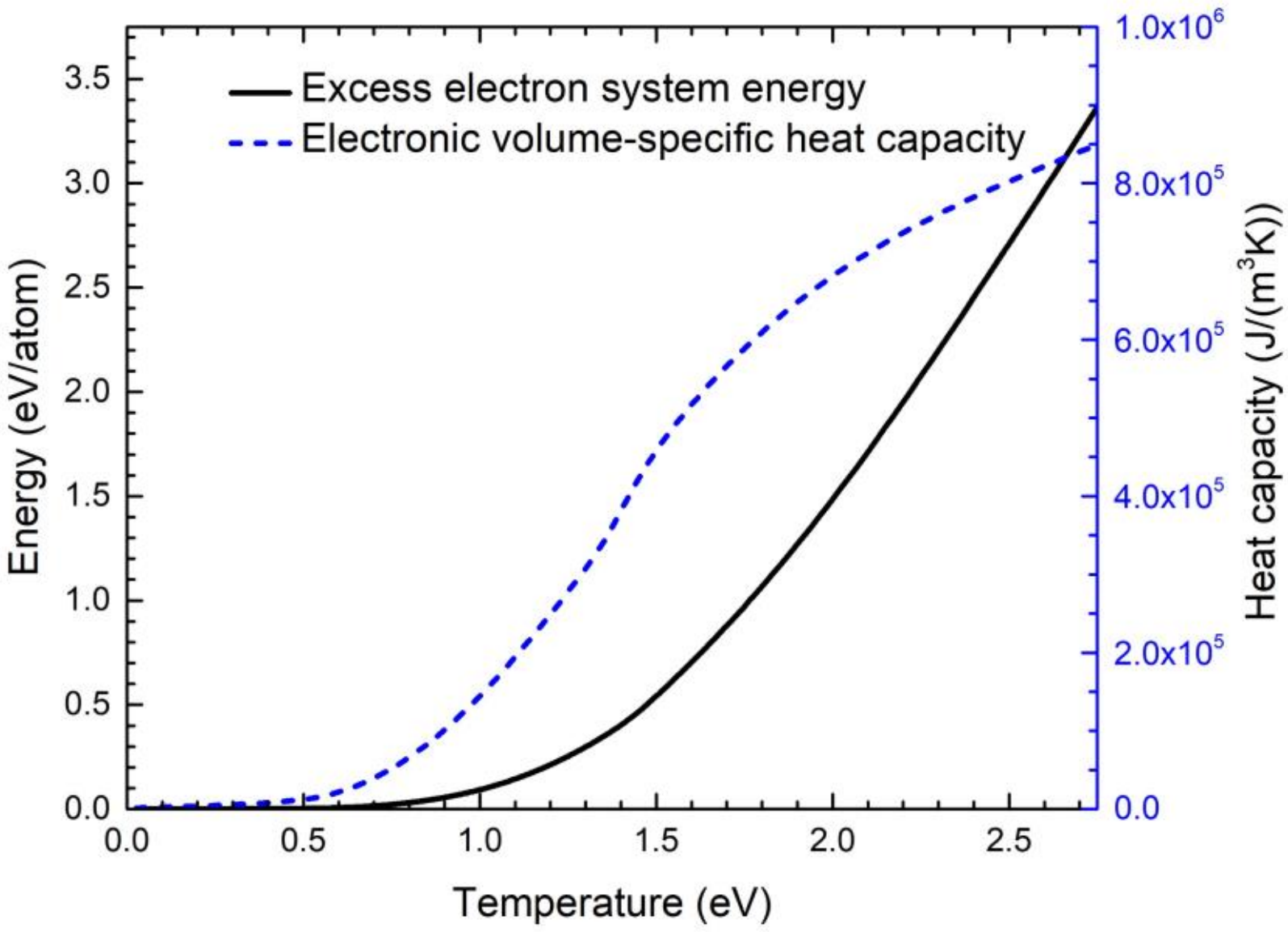

**Figure 1.** Dependence of excess energy of the electron ensemble on the electronic temperature in $Al_2O_3$, calculated by means of self-consistent DFT (solid line, left axis), and the corresponding electronic heat capacity (dashed line, right axis).

For each combination of atomic and electronic kinetic temperatures, we perform DFT-based molecular dynamics simulations to trace evolution of the band structure in alumina within 100 fs after the projectile impact (the characteristic timescale of electronic relaxation in SHI tracks [30]).

In these simulations, we apply fhi98PP norm-conserving pseudopotentials from the ABINIT library [31] and the Perdew-Burke-Ernzerhof (PBE) exchange–correlation functional [32]. Although non-hybrid functionals are known to underestimate the band gap value at ambient conditions, in the case of high electronic temperatures they perform much better [33]. The simulated $Al_2O_3$ supercell is composed of 2 × 2 × 2 rhombohedral (R-3c) primitive cells with 10 atoms in each (80 atoms total) and the lattice parameters a = 10.054 Å, α = 55.342º. The energy cutoff parameter controlling the size of the plane wave basis set used is $E_{cut} \approx 1088$ eV (80 Ry). During DFT-MD simulations, the gamma point is used for calculations of forces acting on atoms, which is sufficient for simulation boxes of our sizes [34,35].

Applying the given average temperature in each cylindric layer, we use an NVT-ensemble (constant number of particles, volume, and temperature) for the electronic system and an NVE-ensemble (constant number of particles, volume, and energy) for the atomic system. The choice of constant volume corresponds to the conditions achieved after SHI irradiation, while changes in the material density require much longer timescales. For all DFT-MD simulations, the time step of 0.25 fs is used.

In all further estimates, we neglect the atomic thermal diffusion between the cylindrical layers during the first 100 fs since the excess energy of the atomic lattice accumulated in the region of a 1 nm radius around the ion trajectory in alumina decreases only by 10-15% within these times (see an example in Figure 2). To estimate the heat diffusion in the atomic system from the central region of the track and model structure changes after SHI irradiation at longer timescales (up to 100 ps), we use classical MD (LAMMPS code [36]) with NVE ensemble and simulation step of 1 fs in an $Al_2O_3$ supercell of the size $22 \times 22 \times 40$ nm$^3$ with periodic boundary conditions. The Berendsen thermostat is applied at the 5 Å thick cell boundaries parallel to the ion trajectory with a damping time of 100 fs [37]. We use the Vashishta interatomic potential [29], which was previously employed to describe SHI tracks and showed good agreement with the experimental data [38].

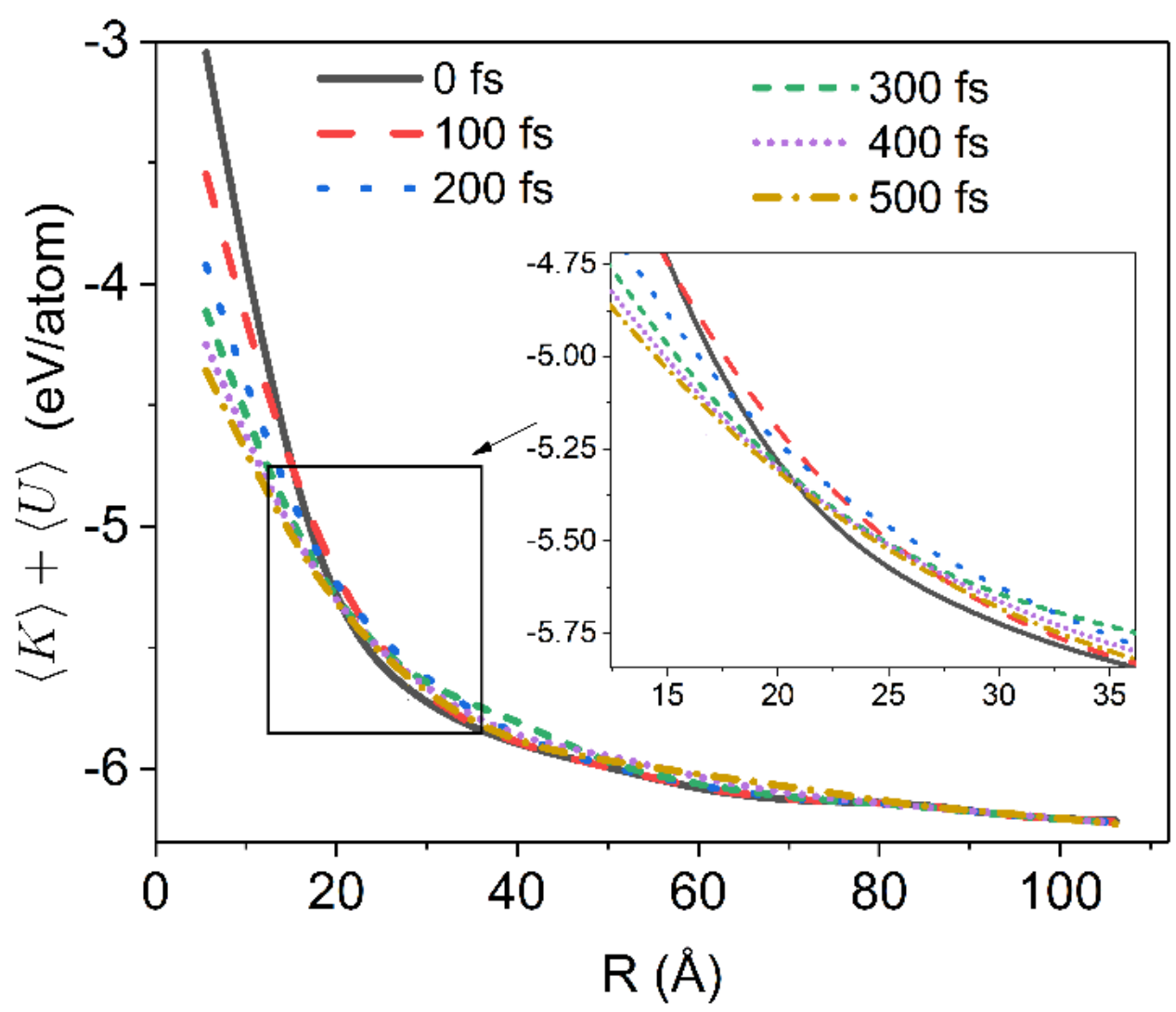

**Figure 2.** An example of an MD simulation of the temporal evolution of the radial distribution of total (kinetic and potential) energy in the excited atomic system of a 23 MeV Fe ion in $Al_2O_3$.

## Results

We simulate alumina under impact of a 700 MeV Bi ion. Figure 3 shows the temporal evolution of the radial profiles of the excess energy accumulated in the atomic lattice (a) and the electronic system (b). The material around the ion trajectory is subsequently divided into cylindrical shells, as shown in the Figure. In each shell, we calculate the average kinetic temperatures of the atomic and electronic systems within the characteristic relaxation time of the electronic excitation (100 fs after the ion impact).

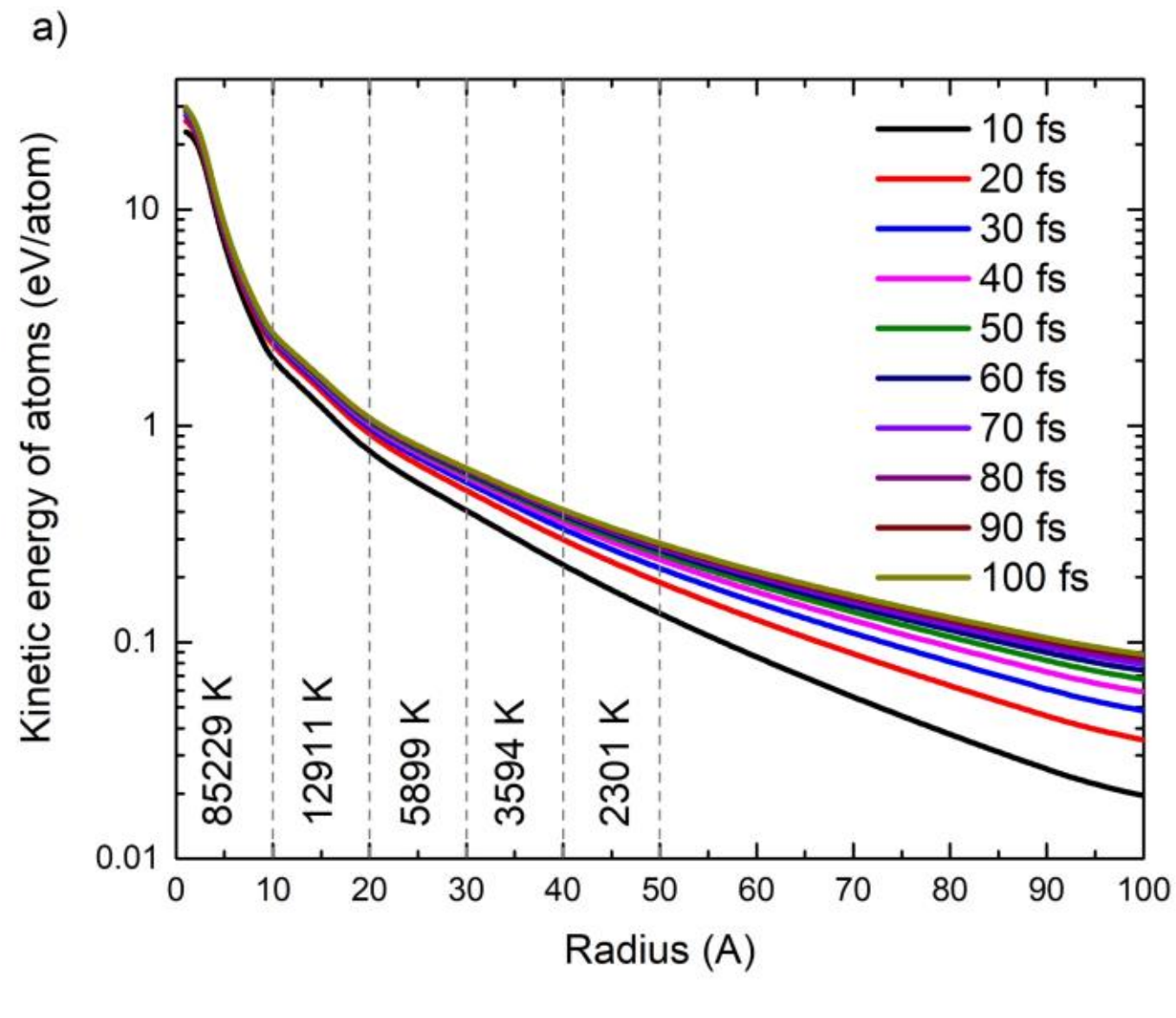

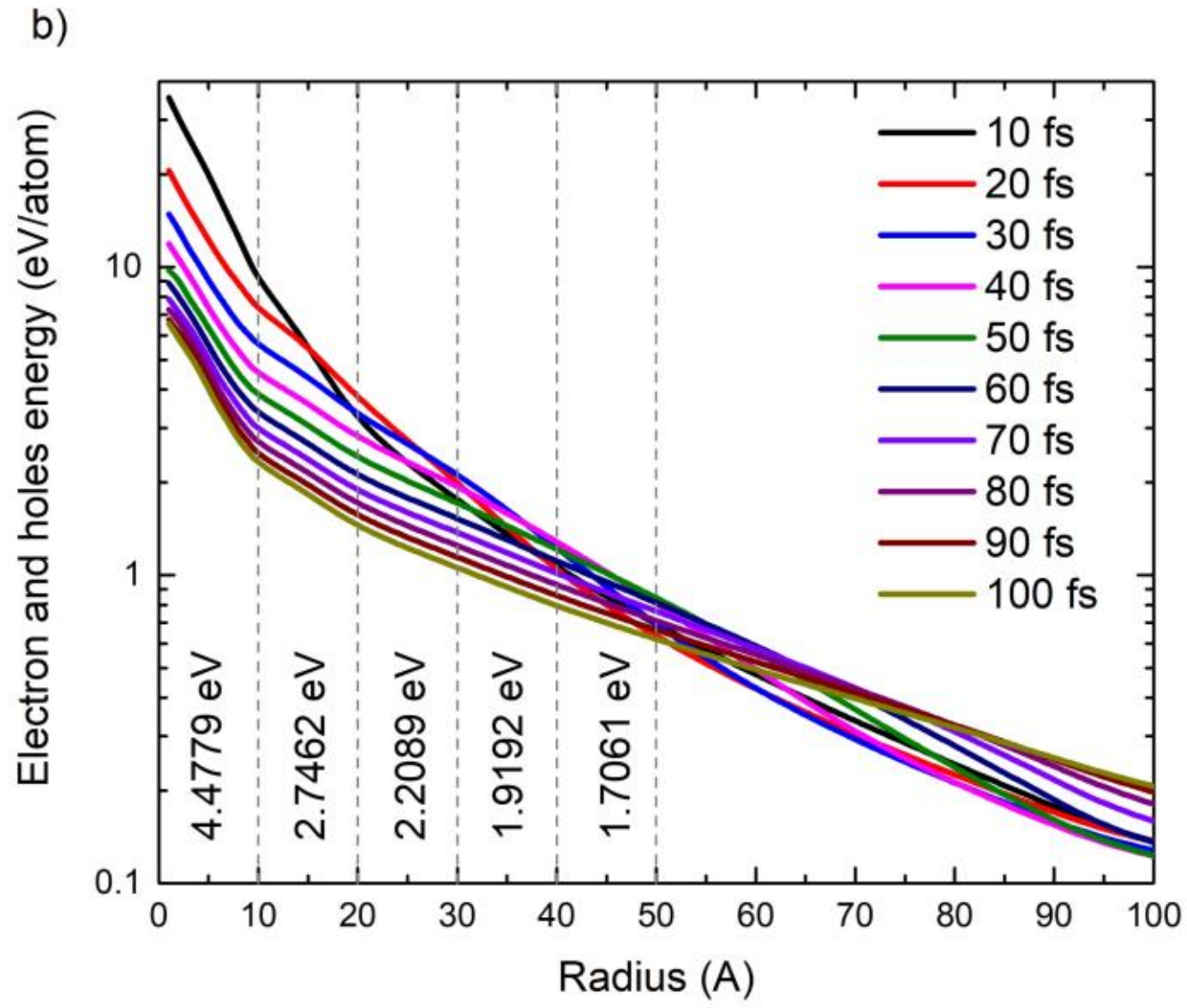

**Figure 3.** Radial energy distributions in $Al_2O_3$ after 700 MeV Bi ion passage. (a) Kinetic energy transferred to the atomic system due to scattering of electrons and valence holes within 100 fs after ion impact, calculated with TREKIS-3. The lowest line is for 10 fs, the highest is for 100 fs. (b) Kinetic and potential energies accumulated in electron and hole systems. Vertical grey dashed lines denote radial layers in which average temperatures are calculated (numbers shown).

Using this set of temperatures, we perform a series of DFT-MD calculations tracing band structure changes over time. Figure 4 illustrates evolution of the electronic energy levels in $Al_2O_3$ after an impact of 700 MeV Bi ion. At characteristic temperatures within ~2 nm radii from the ion trajectory, the band gap collapse occurs within 20 fs. The considerable band gap shrinkage takes place at temperatures realizing at ~2-4 nm radii from the ion trajectory and the band gap is almost unaffected at lower temperatures (larger radii).

These results clearly demonstrate transient spatial inhomogeneity of the band structure in the excited track. The gradients of the deposited energy produce a gradient in the band gap value around the SHI trajectory.

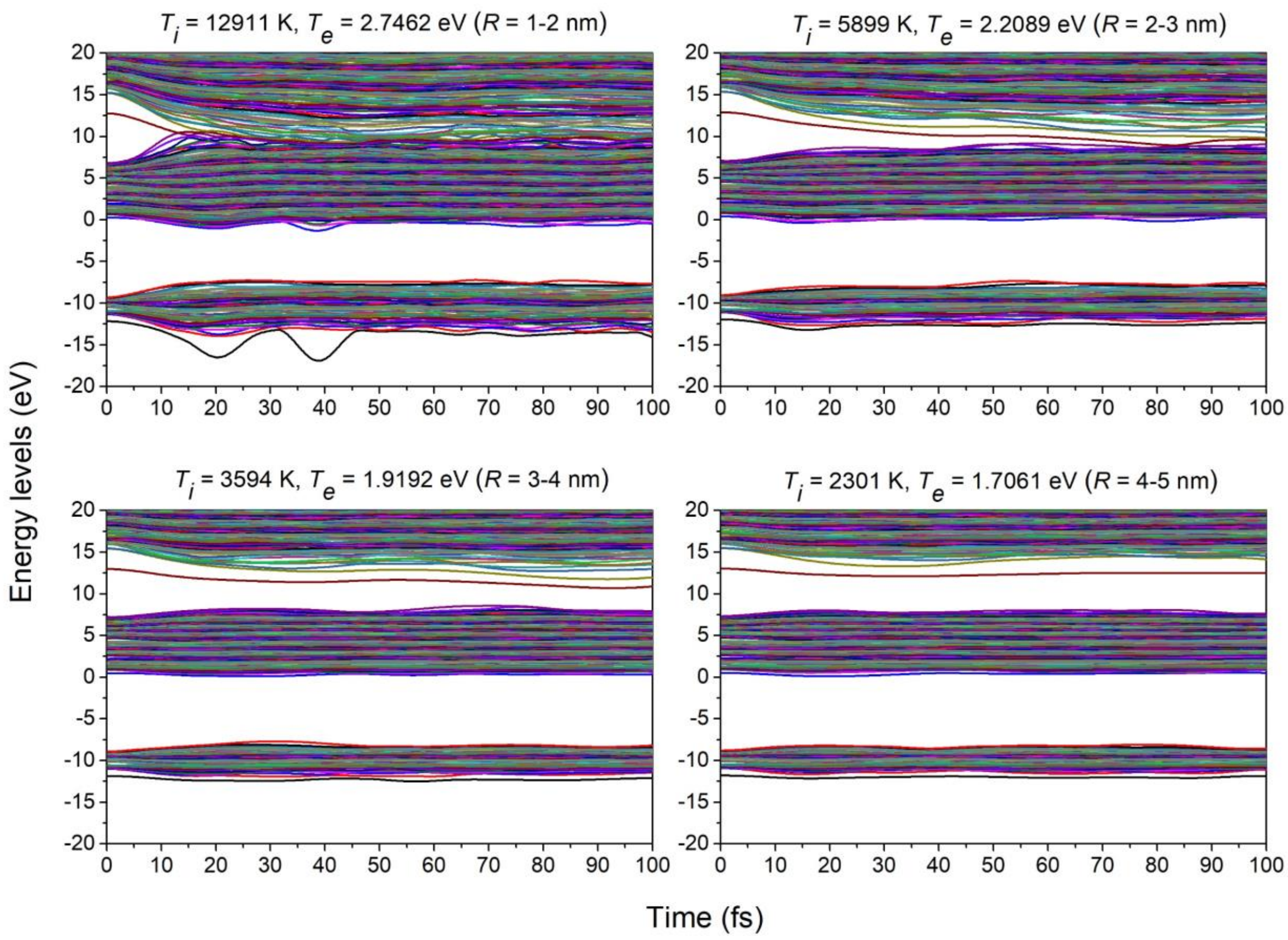

**Figure 4.** Changes in the electronic energy levels in $Al_2O_3$ caused by nonthermal atomic displacements in different radial layers in a track of 700 MeV Bi ion.

The spatial inhomogeneity of the band gap width creates a transient state resembling a metal-semiconductor heterojunction (see schematics in Figure 5). As was discussed in the literature, spatial variations of the band gap may affect low-energy electrons transport [39]. In our case, it may hinder their diffusion outwards from the track center in the heterogeneously modified conduction band, affecting the energy distribution, with more energy stored at the center of the track and less at the periphery (cf. Figure 3b). Moreover, this locking of electrons and holes due to the gradient in the chemical potential may further increase the amount of energy transferred to the atomic lattice in the track core.

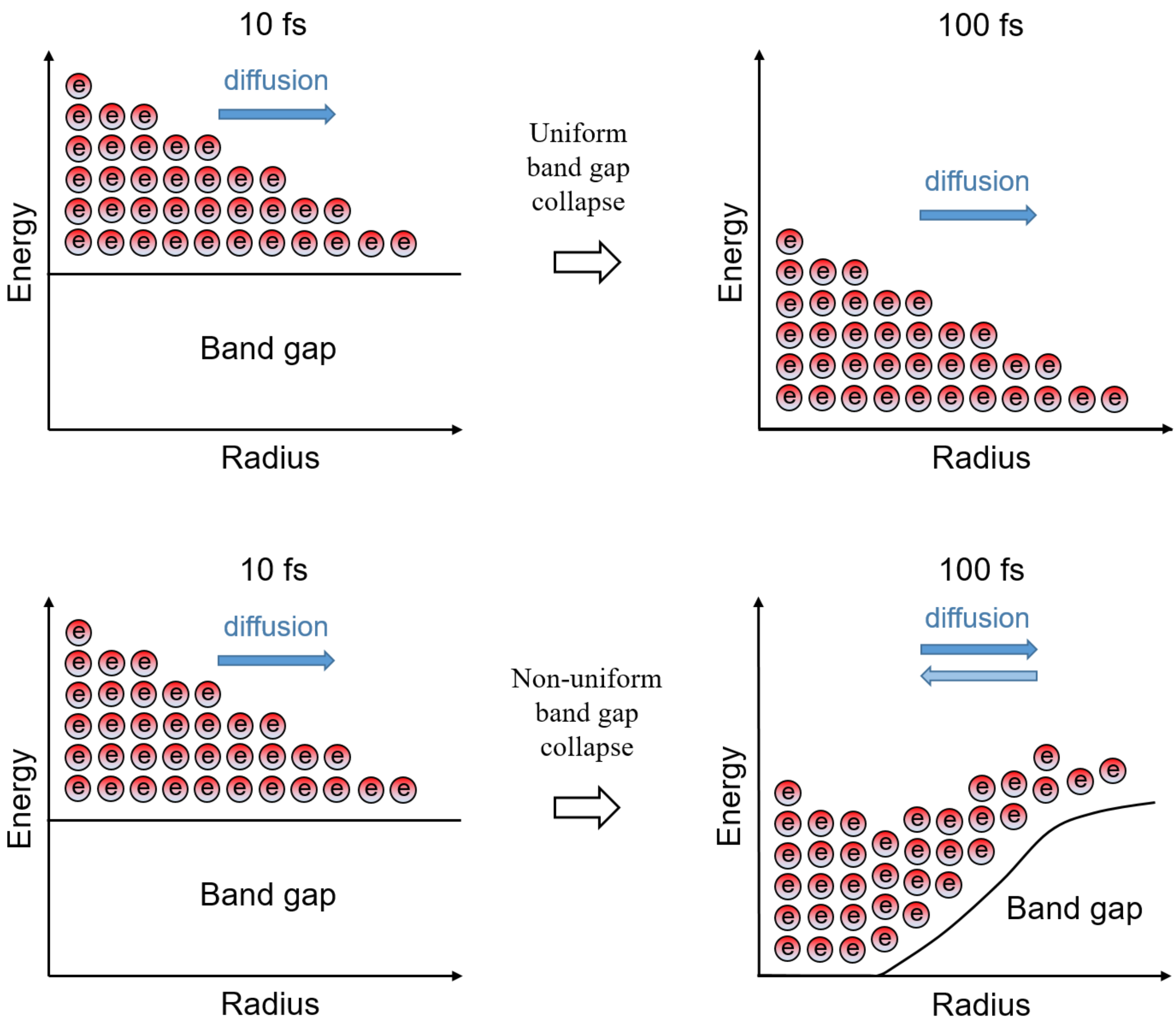

**Figure 5.** Schematics of transient metal-insulator-like heterojunction in SHI track in $Al_2O_3$.

Bandgap narrowing also alters the potential energy of electron–hole pairs generated in the ion track. An incomplete collapse of the band gap leads to only partial transfer of the initial potential energy to the atomic system via the nonthermal energy transfer mechanism. We estimate this effect on the creation of a damaged SHI track using molecular dynamics simulation of the atomic response to the excitation, approximating the transformation of the potential energy of electron-hole pairs into the kinetic energy of atoms according to the extent of the band gap reduction (extracted from Fig.4) at different distances from the ion trajectory, see Figure 6.

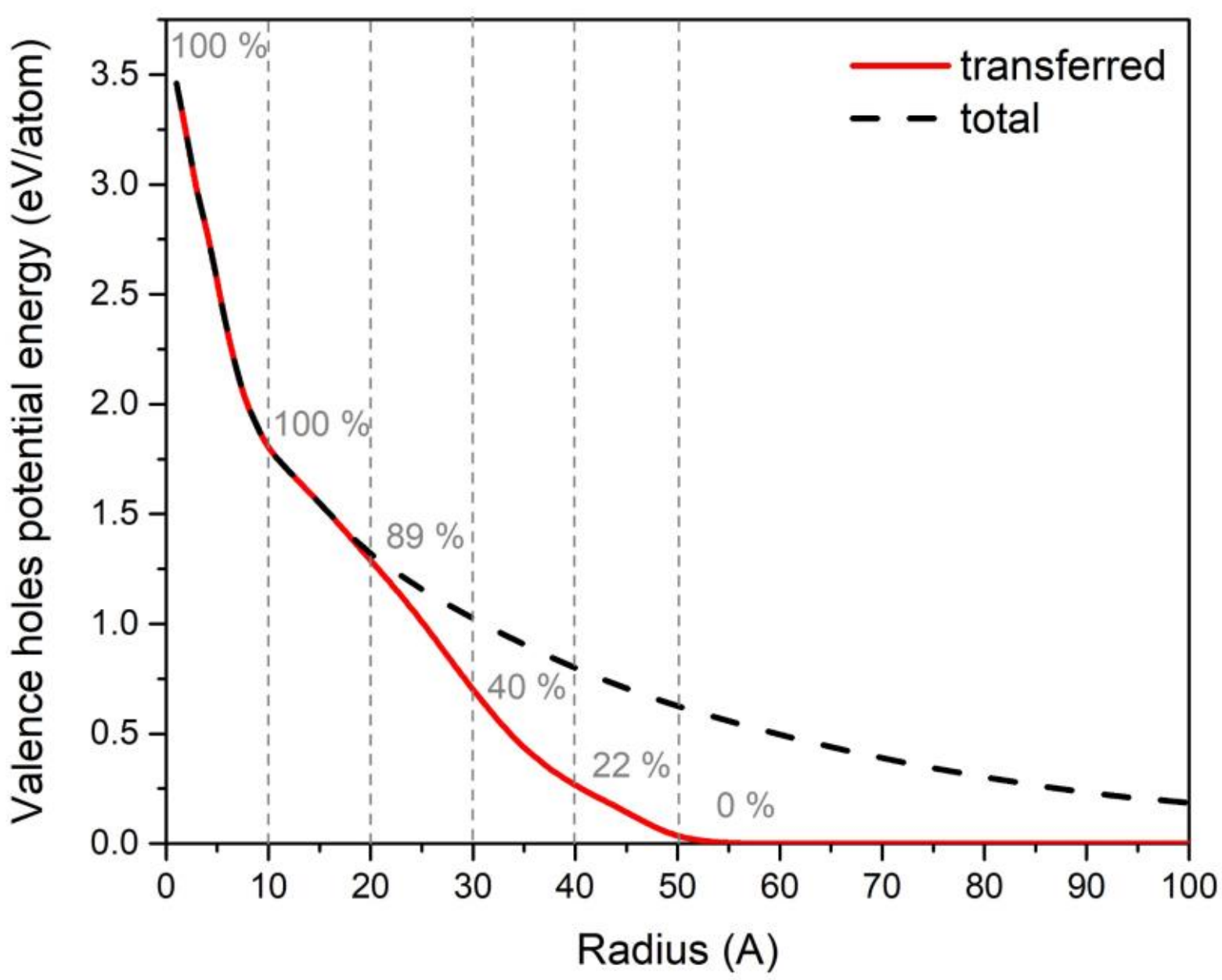

**Figure 6** Radial distribution of the electrons-valence holes' potential energy, delivered to atoms in around the trajectory of a 700 MeV Bi ion in alumina due to partial band gap collapse. A portion of the transferred energy is shown with percents in each cylindrical layer.

Figure 7 shows the results of MD simulations of damage of alumina by a 700 MeV Bi ion, when heterogeneous transfer of the potential energy of electrons and valence holes is taken into account (according to Fig. 6). It is compared to the previously used approximation, in which all the potential energy is delivered to the atoms (cf. Figure 8). In Figures 7 and 8, the X, Y, and Z axes coincide with the crystallographic directions $[\bar{1}2\bar{1}0]$, $[\bar{1}010]$, and $[0001]$, respectively.

The track diameters in both simulations are reasonably close to the experimental data: 3.4 nm for the inhomogeneous band gap collapse (Figure 7) and 4.1 nm when the transfer of the full potential energy of electron–hole pairs to the atomic lattice is assumed (Figure 8), compared to the experimental value of 3.5±0.5 nm [20]. This is, in part, a consequence of strong recrystallization of the primarily damaged track region in $Al_2O_3$ [10]. Also note that according to Ref. [40], the visibility of Al atoms occupying vacant octahedral positions is significantly reduced, making those defects invisible from certain directions.

However, upon a closer examination, the track structure along the X and Y axes in Figs. 7 and 8 differs: the cruder approximation in Fig. 8 shows radial track size marginally outside of the experimental error bars and a much wider defect halo vs. more precise results in Fig. 7. The more advanced simulation with the partial band gap collapse (Fig. 7) reproduces better the measured track diameter, and also demonstrates a more pronounced discontinuous nature of the damage along the ion path, the same as detected in the experiment [16].

Taking all this into account, we may conclude that a careful evaluation of the nonthermal effects and associated changes in the electronic structure are essential to reproduce fine effects of structural modification in SHI tracks in insulators.

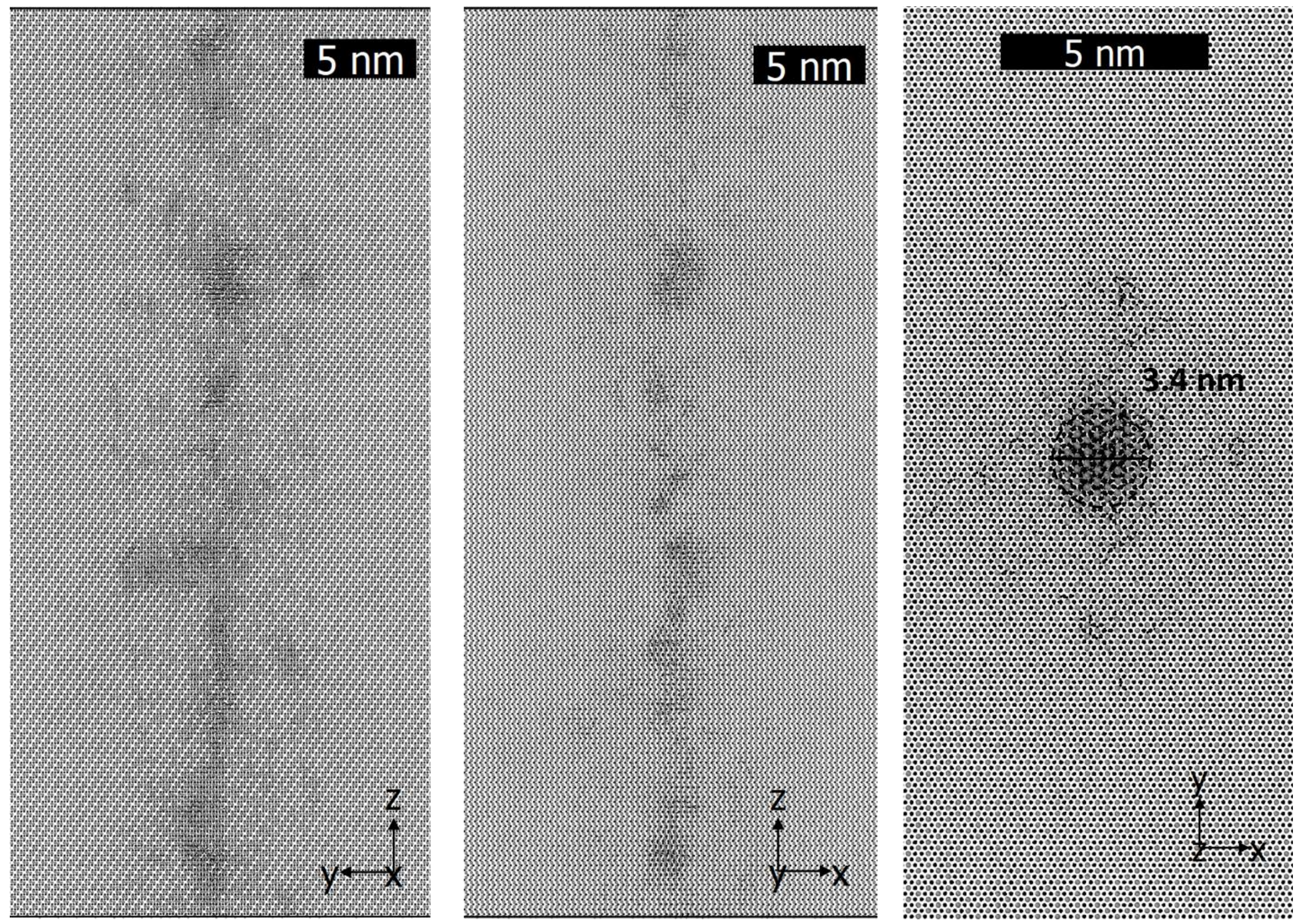

**Figure 7.** Atomic snapshots in $Al_2O_3$ along different orientations at 100 ps after the passage of a 700 MeV Bi ion when partial transfer of the potential energy of electron-valence hole pairs is assumed.

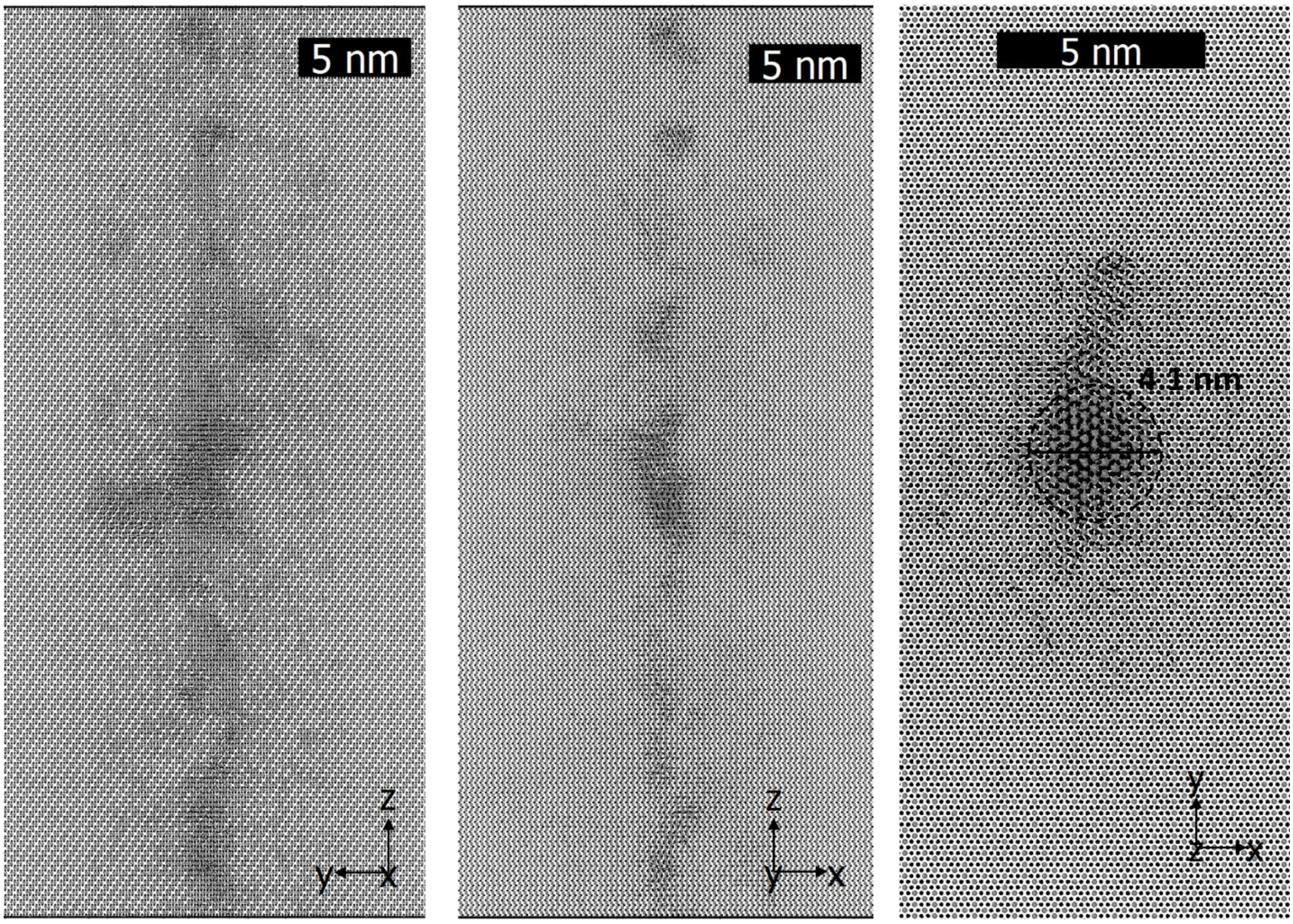

**Figure 8.** Atomic snapshots in $Al_2O_3$ along different orientations at 100 ps after the passage of a 700 MeV Bi ion when the whole potential energy of electron-valence hole pairs is transferred to atoms.

## Conclusions

In this work, we used a combination of Monte-Carlo model and DFT-MD simulations to study the effect of the radial inhomogeneity of the nonthermal energy transfer to the atomic system around the ion trajectory on the formation of damage in swift-heavy ion tracks in $Al_2O_3$. Simulations of 700 MeV Bi ion impact showed that the nonthermal band gap collapse, essential for fast transfer of the excess potential energy of electron-holes pair to the atomic lattice, occurs only in the central area of an SHI track in alumina within ~2 nm radius from the ion trajectory, diminishing with the distance outwards from the track core. Transferring this energy to the lattice according to the degree of band gap shrinkage improves the results on damage simulation. Accounting for this effect enabled us to reproduce both, the track diameter and fine discontinuous SHI track structure in $Al_2O_3$.

**Acknowledgements**

NM gratefully acknowledges the financial support from the European Commission Horizon MSCA-SE Project MAMBA [HORIZON-MSCA-SE-2022 GAN 101131245], and from the Czech Ministry of Education, Youth, and Sports (grant nr. LM2023068). This work was carried out using computing resources of the federal collective usage center Complex for Simulation and Data Processing for Mega-science Facilities at NRC "Kurchatov Institute" (http://ckp.nrcki.ru/).